*Evgeny Bezlepkin*

# PHILOSOPHICAL FOUNDATIONS OF INTERPRETATIONS OF QUANTUM MECHANICS

e-mail: evgeny-bezlepkin@mail.ru

**Abstract** It is demonstrated that the reason for the diversity of interpretations of quantum mechanics is that they are not connected by continuity relations with classical physics, and also the reason is the impossibility of operationalist definition of the "vector of state". The problem lies in the incompatibility of the philosophical foundations of interpretations, which results in the difficulty of building a unified picture of the world. To solve the problem, we identify general philosophical foundation of interpretations of quantum mechanics and built their classification. We also show that in more general theories, the part of which is quantum mechanics, it is possible to integrate (reconcile) the philosophical foundations of interpretations.

## Introduction

There are many interpretations of quantum mechanics, the foundations of which seem to be inconsistent (for example, deterministic and indeterministic interpretations). In our opinion, it is this incompatibility that is the philosophical problem of the diversity of interpretations.

In connection with this, the objective of this article is, first, to identify the cause of plurality of interpretations; second, to distinguish the philosophical foundations for the classification of interpretations and create such a classification; and, third, since the incompatibility of the foundations of interpretations presents a problem for building a unified picture of the world, it is necessary to demonstrate a possibility of minimizing the number of interpretations, if we require their agreement with generalized theories such as, for example, the string theory.

A unique ontology (interpretation) of quantum mechanics (QM) cannot be built solely on its own basis. To do this, one should use a holistic approach; i.e., one should reconcile QM with the theories, which are connected with other areas of applicability. Roughly speaking, plurality of interpretations is due to the fact that quantum theory does not "resist" free interpretation of its basic concepts. It is possible that if quantum mechanics existed as a part of another, more general theory, the latter would impose tighter restrictions on the interpretation of the basic concepts of QM. Thus, it seems that it is possible to describe the world in an integral and consistent fashion through a unified theory, to which there will corresponds a unified ontology, whereas the ontology of quantum mechanics will become one of its parts.

**The main ideas of quantum mechanics**

In order to indicate the source of plurality of interpretations of quantum mechanics, we should clearly identify its basic premises.

For these purposes, we use the theory (formalism) of P. Dirac, because it is the most general one. The initial position of the theory is that to the variety of physical concepts there correspond mathematical concepts (this correspondence is called "parameterization" or "coordinatization"). Thus, the physical objects, which are to be mathematically (quantitatively) studied, should be "parameterized" or "measured". B. G. Kuznetsov points out that this premise, together with the idea of invariance, is the basis of Dirac's theory. He writes that in QM there appears "independence of the physical processes from the methods of observation, expressed in the invariance of certain physical quantities with respect to transition from one coordinate parameterization to another" [Kuznetsov 1957, p. 211].

The first thesis of Dirac is the principle of superposition of states, which says that any individual state of a quantum-mechanical system can be expressed in terms of a linear superposition (overlay) of several other states [See: Dirac 1979, p. 24].

The second point is that to the physical concept of "the state of a dynamic system" there corresponds the mathematical concept of "vector" (in an infinite-dimensional Hilbert space) [Ibid. pp. 29-34]. A fundamental quantity in this correspondence is not the "length" of vector, but its "direction" in the vector space, because in this space the Archimedes' axiom concerning the comparison of lengths does no work; that is, in the space of vectors of quantum mechanics it is impossible to determine which vector is "larger" and which is "smaller".

Dirac gives the vector of state the name "ket-vector", and, having in mind that to any given collection of vectors one can associate a set of dual vectors, he introduces a concept of "bra-vector". These vectors describe the state as well as the superposition connections of a quantum-mechanical system.

The third thesis. While in classical physics the state of a physical system is described by dynamical variables (coordinate, momentum), in quantum mechanics to the dynamic variables, describing a physical quantity, there correspond linear operators. In addition, the algebraic commutative law of multiplication does not hold for them (AB ≠ BA) [Ibid. p. 40].

The fourth point is formulated as follows: "In whatever state a dynamic system is, the result of measuring a real dynamic variable is one of the eigenvalues of this variable" [Ibid. p. 54]. The measurement process, performed on a quantum-mechanical system, causes a disturbance which results in a "jump" of the system from an indeterminate (probabilistic) state to some definite state (exact in the numerical sense). This assertion has been called the projection postulate. Thus, each observable quantum-mechanical quantity has a definite value only in a state which is called an eigenstate; in other states, it has only a statistically average value.

The final thesis of Dirac is that any quantum-mechanical system is described by a complete set of compatible dynamic variables, the observables (e.g., momentum, coordinate, spin, strangeness, etc.). Note that there is no experimental criterion that can confirm the completeness of a given set of variables.

In addition, let us note the thesis about the probabilistic interpretation of the "vector of state" concept, which is that the probability of the measurement result of a quantum-mechanical variable equals the squared modulus of the state vector (the wave function) of this variable. The wave function determines the probability distribution of all the observed quantities, which correspond to a definite state of the system. If the wave function is known, then we know the probability distribution of the observed quantities. Thus, if we repeat the experiment a large number of times, a specific result is obtained in a fraction of the total number of trials, therefore we say that there is a probability of getting a certain result.

In our opinion, the above theses are sufficient for conceptual characteristics of Dirac's theory.

**The concept of the vector of state**

The main object of quantum mechanics in Dirac's theory is the notion of "the state of a quantum-mechanical system," to which there corresponds the mathematical concept of the "vector of state". However, "the mathematical formalism of quantum theory can be constructed differently and, as a starting point, there can be taken any other of the indicated objects [vector of state, operator, invariance group, the probability amplitude and so on. – E.B.], whereas all the other objects will be defined through it" [Sudbury, 1989, p. 277]. Note that a free choice of the base object is inherent to any axiomatics. In this case we get not the alternative interpretations, but the alternative formalisms of QM (e.g., the algebraic formalism, which is based on the concept of operator).

In this regard, we point out the differences between the theories (formalisms) due to Heisenberg and Schrödinger. Dirac conjectured that any system of particles can be characterized by two variable (in time or in space) quantities: either by the state of the system, or by dynamic variables (operators).

The Schrödinger equation is derived from the assumption that "the change of dynamic conditions over time can be attributed to … the change in the state, while considering the dynamic variables constant" [Dirac 1979, p. 153]. The basic concept here is the wave function. The Heisenberg equation is derived from the assumption that the dynamic variables change, while the state is considered invariable. The basic concept here is the linear operator.

Thus, comparing the considered formalisms, we can say that it is in the Dirac theory where there arises independence of "non-classical object from the choice of ways of the classical (conditionally classical) description. This independence is manifested in the existence of a certain transformation law, certain transformations, taking one classical representation of a non-classical object to another "[Kuznetsov 1957, p. 217].

However, since the vector of state (and the principle of superposition) is one of the basic concepts of QM, other formalisms do not exclude it, so it is still possible to apply to them all the existing interpretations. Note that interpretations cannot be differentiated empirically since they do not lead to distinguishing experimental consequences. If such a distinction existed, different interpretations would be considered different theories.

Starting from P. Dirac's postulates and in agreement with A. Sudbury, we will argue that an "interpretation of quantum mechanics is essentially an answer to the question 'What is the vector of state?' "[Ibid. p. 292]. In Schrodinger's formulation of quantum mechanics, there corresponds to it the concept of the wave function; in Heisenberg's formulation there corresponds to it a certain linear operator.

As it was stated above, the correspondence between the concept of "vector of state" and the concept of "the state of a quantum-mechanical system" is not one-to-one (the vector of state is not changed after multiplication by a complex number). Because of this, we conclude that the *vector of state is not defined in the operationalist sense*.

In this regard, the following questions arise. Firstly, whether the plurality of interpretations is due to the fact that the state vector is not defined operationally? Secondly, is it possible to operationally define the state vector (does the parametric method of constructing the quantum theory impede this)?

To answer the first question, we compare the concept of the vector of state with the main object of Maxwell's theory of electromagnetism, the intensity vector of electromagnetic field.

Electromagnetic field (for concreteness, let us consider electrostatics, in which only the electric field appears) is characterized at each point of the space by the magnitude and direction of the force, so the electric field can be characterized by the electric field vector. Electric intensity is "the force which would act on a small body, charged with a unit positive charge, if it were placed at that point without distorting the existing distribution of electricity" [Maxwell, 1989, p. 86].

This definition has two specific features which are decisive for us. First, Maxwell immediately interprets his concept. He writes that "electric intensity is a force", and this interpretation creates a relation of continuity with classical mechanics. Secondly, after the interpretation there comes a description of a specific set of actions (the ideal experimental procedures), by doing which we obtain the magnitude and direction of the electric field vector.

Turning to consideration of quantum mechanics and the concept of the state vector, we can note that, firstly, neither Schroedinger, nor Dirac interpreted their key concepts. For example, Schrödinger wrote: "It is quite natural to associate the function $\psi$ with some oscillatory process in the atom ... Until more complicated problems are calculated by the new method, I do not think it possible to consider in more detail the interpretation of the introduced oscillatory process" [Schrödinger 1976, p. 18]. Second, the state vector corresponds to the physical system not in a one-to-one fashion. That is, by virtue of the specific mathematical nature of this concept, there does not exist an experimental procedure to distinguish the lengths of one and the same vector of state. If there is a quantum particle or a system, then it is possible to associate with it a set of equally directed vectors of state. Third, the state vector can be considered, at least, from three points of view: the ones due to Schrodinger, Heisenberg and Dirac (the combining one).

Thus, we conclude that the existence of multiple interpretations of quantum mechanics is connected with the following reasons:
1) no author's interpretation of the basic concepts of QM, consistent with classical physics, is given;
2) in most cases, the external interpretations do not have continuity relationships with classical physics;
3) it is impossible to define the concept of "the state vector" operationally; i.e., in the form of a collection of experimental procedures that provide the rules for distinguishing of the state vectors. The impossibility of an operationalist definition is inherent in the axioms of quantum mechanics.

**Specific features of quantum mechanics**

None of the interpretations of QM can express the real ontology of the world, because they are just some possible interpretations. Because of their multiplicity, a consistency check is required. To this end, we will first examine in more detail the "composition" of quantum theory.

QM consists of two blocks: a theory of a closed quantum system (the system is described by the Schrodinger equation which deterministically characterizes its behavior prior to the measurement process); and a theory of measurement (the projection postulate describes interaction of the system with a measuring device (an act of measurement), and it has a fundamentally probabilistic nature).

Thus, speaking about the features of QM, we are talking mainly about the features of the theory of measurement. Here are the main features:
1) indeterminacy (there do not exist simultaneous definite values for those characteristics which a particle is endowed with in classical mechanics; the status of observables, when the system is not in an eigenstate, is not defined).
2) inseparability (the possibility of obtaining information about one of quantum-mechanical subsystems by performing experiments with another).
3) the projection postulate. It introduces: duality (subdivision of the world into a microscopic object and a macroscopic measuring device); indeterminism (the measurement results are probabilistic in nature); also it does not take into consideration the possibility of continuous observations.

In addition, the use of the projection postulate leads to the notion of collapse of the wave function. For correct understanding of the collapse problem, one should subdivide the measurement in QM into three stages: preparation of the initial state, perturbation, determination of probability of the final state. Now we present several views on this phenomenon.

Let us quote a statement by A.I. Markov about the "screen with a slit" experiment: "Screen with a slit can perform various functions. In the region of preparation it will serve as a "classic" filter, which prepares the original state ... it is included in the technical operations and is outside the region of applicability of the language of wave functions ... Only being inside the studied system, the screen with a slit will be a 'quantum-mechanical' filter described by the projection operators introduced by von Neumann and Dirac"[Lipkin 2010, p. 83]. N. Bohr provided the same example in his discussions with Einstein, so the above-said can be included into the Copenhagen interpretation of quantum mechanics.

Here is the statement by V.A. Fock about the same experiment. "In the performed experiment there has been realized one of potential outcomes provided by the initial wave function. The change in the formulation of the question about probabilities is exactly in taking into account the realized result, i. e., in taking into consideration new data. In addition, to new data there corresponds a new wave function " [Fock 1957, p. 472]. On the basis of this statement it is possible to develop an interpretation of QM, which is now called the propensitive interpretation. Thus, one can find such interpretation which either exclude or logically explain the emerging collapse of the wave function.

These statements will be used by us for further distinguishing of foundations and classification of interpretations of QM.

**Analysis and classification of the interpretations of QM**

Let us proceed to the analysis of interpretations of QM. On the basis of analysis of the statistical, Copenhagen and many-worlds interpretations of QM, M.A. Markov identified the following positions which these interpretations share ([Markov 1991, pp. 70-100]).
1) Requirement of carrying out normalization for the finite motions.
2) Existence of randomness as absolute chance.
3) Adoption of Bohr's propositions: for the interpretation of measurements it is necessary to use the classical ideas; all experimental data must be described with the help of classical concepts.
4) Adoption of the existing equations of quantum mechanics.

From the above statements we can derived a minimal interpretation of QM (a sort of Copenhagen interpretation). It should be noted that these statements do not touch on the concepts of the state vector and the projection postulate. Therefore, in accordance with M.A. Markov and A. Sudbury, we describe interpretations of QM on the basis of these two criteria: the interpretation of the state vector and the interpretation of the projection postulate. According to Dirac, the vector of state is the primary concept of quantum mechanics. All that one can say is that the concept of the state of a physical system is algebraically reduced to the mathematical concept of vector in an infinite-dimensional space. Thus, "to each state of a dynamical system at any given moment of time there corresponds a ket-vector" [Dirac 1979, p. 29]. The projection postulate is used for the moment of measuring the system. In this regard, Dirac states that "a measurement always causes a jump of the system into an eigenstate of that dynamic variable the measurement of which is performed" [Ibid. p. 54].

Here is a list of interpretations according to [Markov 1991, pp. 70-100], [Pechenkin 1999], [Sudbury 1989, pp. 288-307], [Sevalnikov 2009, pp. 69-76].

Table – Characteristics of interpretations of QM

| interpretations | vector of state (VS), projection postulate (PP) |
|---|---|
| 1) Copehagen (minimal) | VS is a mathematical method for calculating the results of experiments; PP is not needed, a distinction is introduced between the procedure of preparation and measurement of the system; each measurement relates to a given preparation, not to that one which will be after the measurement. |
| 2) objective (literal) | VS is an objective property of the system; it lies in one of subspaces of the space of states, the system switches between subspaces, the probabilities of which are determined by solving the Schrodinger equation; PP says about the changes of VS after measurement, so indeterminism is a property of the world. |
| 3) Copehagen (standard) | VS is a list of possibilities (superposition of states), one of which is realized during the interaction between a micro-object and a measuring device (the collapse of VS, indeterminism). The standard interpretation of PP. |
| 4) subjective | VS reflects the level of experimenter's knowledge about the system. VS changes after measurement, because the level of experimenter's knowledge about the system has changed. |
| 5) ensemble | VS describes not a single particle, but an ensemble of particles. The squared modulus of VS describes the proportions of systems in the ensemble, for which a certain result was obtained. These parts are subsystems of the ensemble; PP describes shifting of attention from one ensemble to another (preserving determinism). |
| 6) many-worlds | VS of any system is defined in relation to the state of the entire universe. A measurement leads to parallelization of the world into many worlds, in each of which a different result is obtained after measurement (preserving determinism). |
| 7) quantum–logical | It claims that all the difficulties of quantum mechanics can be resolved through the use of three-valued non-Boolean logics. |
| | |

| 8) neorealistic (de Broglie - Bohm) | Evolution of a particle is completely determined by a "pilot-wave" (a latent variable) together with the Schrödinger equation. VS depends on the configuration of the entire universe (the postulate of non-locality). PP is not needed, since determinism is recognized. |
| | It is a new theory; however, it can be considered as interpretation, if we assume that it will not make predictions different from the predictions of QM. |
| 9) propensitive, Feynman | VS is a characteristic of potential possibilities of the measurement result. PP describes the transition of a potential possibility into reality (preserving causality). |
| | VS is the probability amplitude, while the full (real) amplitude of probability for a microscopic object is defined as the integral of amplitudes over all possible trajectories of motion. |

We can classify the listed interpretations on the basis of binary oppositions (in particular, based on the article by A.A. Pechenkin [Pechenkin 1999]), whereas these oppositions can be interpreted as philosophical foundations of interpretations of QM.

| Recognition of the real world (interpretation of VS) | subjective **1,3**,4,7 | objective 2,5,**6,8,9** |
|---|---|---|
| Relation to the projection postulate (PP) | dualistic **1**,2,**3**,4 | monistic 5,**6,8,9** |
| Relation to determinism | indeterministic **1**,2,**3** | deterministic 5,**6,8,9** |
| The number of objects described in the system | ensemble 5 | non-ensemble 1,2,3 (others are indifferent) |
| interpretation of a physical quantity in a non-eigenstate | indeterminism **1**,2,**3**,4 | It is exact, but unknown 5,**6,8** (for 9 the value is "fuzzy") |

This table confirms that two issues remain the central ones: interpretation of the state vector and interpretation of the projection postulate. It also confirms that the philosophical foundations of interpretations are contradictory. Among the most obvious contradictions are the following three: subjectivism and objectivism, determinism and indeterminism, dualism and monism.

In this regard, the question arises of whether it is possible to minimize (through coordination, unification or elimination of foundations) the number of different interpretations and to what extent, if it is possible.

To answer this question we would like to bring in a holistic approach, which consists in the fact that a unique ontology (interpretation) of QM cannot be built solely on the basis of QM itself. We think that it is possible to describe the world in an integral and consistent manner through a unified theory, to which there will corresponds a unified ontology, whereas the ontology of quantum mechanics will become one of its parts. That is, we should require coordination of the interpretations of QM with the theories, which are generalizations of quantum mechanics such as quantum field theory and the theory of superstrings. Thus, it is as if we added some "external constraints" on the definition of the "state vector."

Let us turn again to Maxwell's theory of electromagnetism. If we try to geometrize (translate into a geometric language of description) this theory, i.e., make it agree with a model of general relativity theory, then we must interpret the components of electromagnetic field not as forces but as the curvature of a space-time 5-dimensional manifold. If we extend this theory to the quantum domain (within the grand unified theory), then we obtain a concept of "quantized field", where there exist particles on which the forces act.

Thus, one extension ("quantum") of the classic electromagnetism theory confirms the interpretation of the vectors of electromagnetic field as forces. Besides this, in the generalized interpretation there are added "force centers" (particles). Another extension ("geometric") leads to a quite different interpretation. A final interpretation of electromagnetism, we think, will be linked to an even more generalized theory. Nowadays, there are theories that attempt to combine both the above interpretations on the basis of the concept of gauge symmetry (for example, an exceptionally simple theory of everything).

However, let us turn to quantum mechanics. For example, in quantum field theory the formalisms (representations) due to Schrödinger and Heisenberg are not equivalent: to construct it, the Heisenberg representation is used. That is, the descriptions, equivalent in the non-relativistic QM, are not equivalent in the relativistic version of the theory, which is used to describe not only particles, but also fields. We can proceed in the very same fashion with interpretations. In our view, some interpretations have to be sensitive to such extension of the range of applicability of the theory; for example, the indeterministic interpretations may turn out to be inconsistent, if we make it agree with the domain of classical physics; on the contrary, the Feynman interpretation allows such an extension, and we will discuss it later.

**Possibility of minimization of the interpretations of QM**

Let us describe an attempt to build a unified ontology based on variational principles of mechanics, and specifically on the basis of the principle of least action. It is known that using various formulations of this principle, one can build classical mechanics, general relativity theory, quantum theory, as well as the quantum field theory.

Thus, the variational principles can be called generalized principles of physics; and, therefore, Feynman's interpretation, which is based on them, can claim the role of generalized ontology, if we make it agree with the philosophical foundations of classical and relativistic mechanics.

The generalized interpretation of QM through the principle of least action corresponds to the sum of interpretations due to Feynman, Fock and late Heisenberg. It is monistic and synthesizes both indeterminism (probabilistic description on the micro-level) and determinism (causal description on the macro-level), because "variational principles include a synthesis of the continuous and discrete aspects of motion and are an expression of the generalized principle of causality in physics" [Polak 1959, p. 879].

Let us fortify what is said using the words of V. A. Fock: "The principle of causality in a general sense should be understood as a statement about the existence of laws of nature and, in particular, those connected with the general properties of space and time (finite propagation velocity of actions, impossibility of influencing the past). With this understanding, quantum mechanics not only does not contradict to the principle of causality, but it gives it a new expression and expands its application to the probability laws "[Fock 1957, p. 467-468].

The basic notion of this interpretation is a potential possibility. V.A. Fock wrote: "A state of an object described by the wave function is objective in the sense that it represents an objective (independent of the observer) characteristic of the potential possibilities of this or that result of interaction of an atomic object with an instrument. In the same sense, it refers specifically to a given individual object. But this objective state is not yet real, in the sense that, for the object in this state, the indicated potential possibilities are not yet realized. The transition from the potentially possible to the realized, the real takes place at the final stage of experiment" [Ibid. p. 468].

Thus, the vector of state is interpreted as the probability amplitude, whereas the total probability amplitude for a microscopic object is defined as the sum of contributions from all *possible* trajectories of movement. The summation of potential possibilities identifies a trajectory with the maximum probability. In this way there takes place the transition from the (potentially) *possible* to the (realized) *actual*. Therefore, to paraphrase Leibniz, we can say that "we live in the most probable of all possible worlds."

An interpretation, connected with the theses of Feynman and Fock, is also being developed, for example, by V.E. Terekhovich [Terekhovich 2012, p. 112], who calls it "the interference of possibilities". The author writes that, to every possible state of a quantum object, there corresponds the wave function, whereas "the wave functions of all possible states of the object … interfere in accordance with the rule of addition of phases. The resulting condition is characterized by maximum stability in the given conditions and is described by the maximum probability "[Ibid. p. 114]. The resulting state is called real, whereas all others are called possible.

This interpretation, based on variational principles, can be extended to the scope of not only classical mechanics, but the relativity theory. This can be done using the string theory as a generalized theory, which is also based on variational principles of mechanics.

According to V.E. Terekhovich, "if the string theory is correct (there are no proofs of this so far), we can assume that the principle of maximum aging is also an approximation of the method of path integrals. When the scale is increased, the pulsating fabric of multidimensional intricate space-time collapses and is smoothed to 4-dimensional. Because only in it the macro-objects are stable, all possible trajectories in 10-dimensional space are reduced, as a result of interference, to the trajectories in 3-dimensional space. In this case, not only classical mechanics, but also the general relativity theory may become a special case of quantum mechanics" [Terekhovich 2013, p. 86].

Thus, it is possible to obtain a coordinated interpretation not only for quantum mechanics, but for classical mechanics, and through it for the relativity theory.

**Conclusion**

In conclusion let us discuss the following question, the answer to which can only be given by the further development of physics: if the discovery of another generalized principle (among the candidates are the principles of symmetry) is connected with another interpretation, will it be possible to combine the above probabilistic interpretation and the new one? Contradictions between interpretations still signify only one thing − a contradiction in the fundamental principles of the theory or our lack of understanding of the theory.

It is most likely that as the construction of a unified theory proceeds, as new phenomena and concepts are included in it, the ontology will be modified, adjusted to them. In view of this, the interpretations of QM will also be modified and will complement each other.

**Bezlepkin Evgeny** - Junior Researcher, Institute of Philosophy and Law of SB RAS.